\documentclass[conference]{IEEEtran}
\IEEEoverridecommandlockouts
\usepackage{cite}
\usepackage{amsmath,amssymb,amsfonts}
\usepackage{algorithmic}
\usepackage{graphicx}
\usepackage{subcaption}
\usepackage{algorithm}
\usepackage{algorithmic}
\usepackage{textcomp}
\usepackage{dblfloatfix} 
\usepackage{multirow}
\usepackage{hyperref}
\usepackage{array}
\usepackage{hhline}
\usepackage{mdframed}
\usepackage{float}
\usepackage[normalem]{ulem}
\useunder{\uline}{\ul}{}

\usepackage[dvipsnames]{xcolor}
\usepackage[most]{tcolorbox}

\newcommand{\myfigureshrinker}{\vspace{-0.4cm}}

\def\BibTeX{{\rm B\kern-.05em{\sc i\kern-.025em b}\kern-.08em
    T\kern-.1667em\lower.7ex\hbox{E}\kern-.125emX}}
    
\begin{document}

\title{LLMs as Evaluators: A Novel Approach to Evaluate Bug Report Summarization\\
}

\author{
    Abhishek Kumar$^*$, Sonia Haiduc$^\dagger$, Partha Pratim Das$^*$, Partha Pratim Chakrabarti$^*$ \\
    $^*$\textit{Department of Computer Science and Engineering, Indian Institute of Technology, Kharagpur, India} \\
    $^\dagger$\textit{Department of Computer Science, Florida State University, Florida, United States} \\
    E-mail:  abhishek16@kgpian.iitkgp.ac.in, \{ppd, ppchak\}@cse.iitkgp.ac.in, shaiduc@cs.fsu.edu
}




\maketitle

\begin{abstract}

Summarizing software artifacts is an important task that has been thoroughly researched. For evaluating software summarization approaches, human judgment is still the most trusted evaluation. However, it is time-consuming and fatiguing for evaluators, making it challenging to scale and reproduce. Large Language Models (LLMs) have demonstrated remarkable capabilities in various software engineering tasks, motivating us to explore their potential as automatic evaluators for approaches that aim to summarize software artifacts. In this study, we investigate whether LLMs can evaluate bug report summarization effectively. We conducted an experiment in which we presented the same set of bug summarization problems to humans and three LLMs (GPT-4o, LLaMA-3, and Gemini) for evaluation on two tasks: selecting the correct bug report title and bug report summary from a set of options. Our results show that LLMs performed generally well in evaluating bug report summaries, with GPT-4o outperforming the other LLMs. Additionally, both humans and LLMs showed consistent decision-making, but humans experienced fatigue, impacting their accuracy over time. Our results indicate that LLMs demonstrate potential for being considered as automated evaluators for bug report summarization, which could allow scaling up evaluations while reducing human evaluators effort and fatigue.



\end{abstract}


\section{Introduction}

Given the large size of most software systems in use today, software developers need to stay up to date with a lot of information captured in software artifacts, which evolve frequently. Software artifact summarization can help, providing developers with short summaries of artifacts or their changes and reducing the amount of information developers may need to consult and remember. Many approaches have been proposed to summarize software artifacts \cite{zhu2019automatic, panichella2018summarization,zhang2022survey}. When evaluating these summarization approaches, automatic evaluation and human judgement are the two main methods of evaluation. Automatic summary evaluation involves using metrics such as BLEU \cite{bleu}, ROUGE \cite{rouge}, and METEOR \cite{meteor} among others, to evaluate software summarization tasks such as bug report summarization, code generation, code documentation generation, etc. However, these metrics, originally developed for machine translation, are difficult to interpret in a software engineering context and often do not correlate with human judgment\cite{Re_4}, making it difficult to identify real improvements. 

Human evaluation studies provide a more comprehensive understanding of model performance, allowing researchers and practitioners to make more informed decisions \cite{intro_9}. However, human evaluation faces significant challenges that impact its effectiveness \cite{intro_7}: it is time consuming and requires substantial effort from experts, leading to delays and increased costs, especially at scale. Coordinating and training evaluators to ensure uniform understanding of criteria adds complexity and expense, while the limited availability of qualified experts restricts scalability. Additionally, evaluators may experience fatigue and decreased attention span when evaluating long tasks, which can further compromise the accuracy and reliability of the evaluation \cite{intro_8}. 


To address these issues, we propose the use of Large Language Models (LLMs) as evaluators for software summarization tasks. Specifically, in this paper we focus on the task of bug report summarization and aim to answer the question: \textit{Can LLMs accurately evaluate the quality of bug report summaries?} Our idea is motivated by the fact that in recent years, LLMs, particularly those based on Transformers have demonstrated remarkable abilities across a wide range of NLP and SE tasks \cite{intro_1, intro_2,du2024evaluating,khan2022automatic}. If LLMs can generate code \cite{du2024evaluating}, test code \cite{schafer2023empirical}, document code \cite{khan2022automatic}, and generate software summaries \cite{ahmed2024automatic}, it stands to reason that they could potentially also perform the task of evaluating software summaries. In this paper, we perform a study to investigate this possibility. 


In our study, we focused on two specific tasks: evaluating bug titles (short summaries of the bug descriptions) and evaluating bug report summaries (based on bug descriptions and related comments). For each task, we compared the performance of three LLMs, namely GPT-4o \cite{gpt4}, LLaMA-3 \cite{llama3}, and Gemini \cite{gemini}, with human evaluators. Participants were asked to assess the correctness and completeness of bug titles and bug report summaries. We provided the same instructions and evaluation criteria with definitions and examples to both human evaluators and LLMs. The goal was to see if LLMs, when provided with carefully structured tasks and standardized evaluation criteria, are capable of performing summary evaluation tasks typically conducted by humans. Our findings show that LLMs performed well in evaluating bug report summaries, with GPT-4o outperforming other LLMs in most tasks. Both humans and LLMs displayed consistent decision-making aligned with predefined criteria, but humans experienced fatigue over time, affecting their accuracy.

Our key contributions are:
\begin{itemize}
    \item We explore the novel idea of using Large Language Models (LLMs) as evaluators for bug report summarization tasks, proposing an automated alternative to traditional human evaluation methods.
    \item We performed the first study comparing LLMs with human evaluators for evaluating bug report summaries.
    \item Our results reveal the potential of using LLMs for automatically assessing software summaries. We provide a replication package containing our study design and results, to encourage further research and replication (\url{https://bit.ly/3zk7qZr}).
\end{itemize}

\section{Related Work}

To the best of our knowledge, there is no prior work exploring the use of LLMs as evaluators in software summarization tasks. However, several studies have highlighted the limitations of traditional automated and human evaluation methods. 

Stapleton et al. \cite{Re_5} showed a disconnect between BLEU/ROUGE scores and developer comprehension, advocating for metrics that reflect practical utility. Evtikhiev et al. \cite{out_of_bleu} found that these metrics often fail to correlate with human judgment in code generation, leading to misleading conclusions about model effectiveness. Mastropaolo et al. \cite{ICSE} critiqued BLEU and similar metrics for not capturing essential code quality attributes, resulting in discrepancies between metric scores and human assessments. Haque et al. \cite{Re_3} demonstrated that word overlap metrics do not accurately reflect the quality of code summaries, recommending semantic similarity metrics like SentenceBERT instead. Roy et al. \cite{Re_4} revealed that small differences in these metrics do not ensure improvement in summarization quality, emphasizing the need for more robust metrics. Liguori et al. \cite{Re_6_who}  highlighted significant discrepancies between automatic metrics and human assessments, especially in complex code tasks. Finally, Steck et al. \cite{Re_7_netflix} cautioned against using cosine similarity in learned embeddings, suggesting alternatives to improve reliability. 

Regarding human evaluations, Clark et al. \cite{Reh_1} highlight that untrained evaluators often struggle to accurately judge advanced natural language generation models, leading to inconsistent and biased assessments. Similarly, Schuff et al. \cite{Reh_2} discuss the inherent subjectivity and high costs associated with human evaluations, which can result in unreliable and non-reproducible outcomes. Izadi et al. \cite{Reh_3} further emphasize the lack of standardization and scalability issues in human evaluations, pointing out that varied expertise among evaluators and ethical considerations add layers of complexity. Iskender et al. \cite{Reh_4} reveal the necessity for standardized procedures to ensure comparability and reliability in human evaluations, noting that a small number of evaluators can undermine robustness. Lastly, Hamalainen et al. \cite{Reh_5} identify a significant issue where misalignment between problem definitions, methods, and evaluations results in invalid and irreproducible findings. 


\section{Study on Evaluating Bug Summaries}

\begin{figure}[!htbp]
\myfigureshrinker
    \includegraphics[width=0.48\textwidth, height=0.15\textheight]{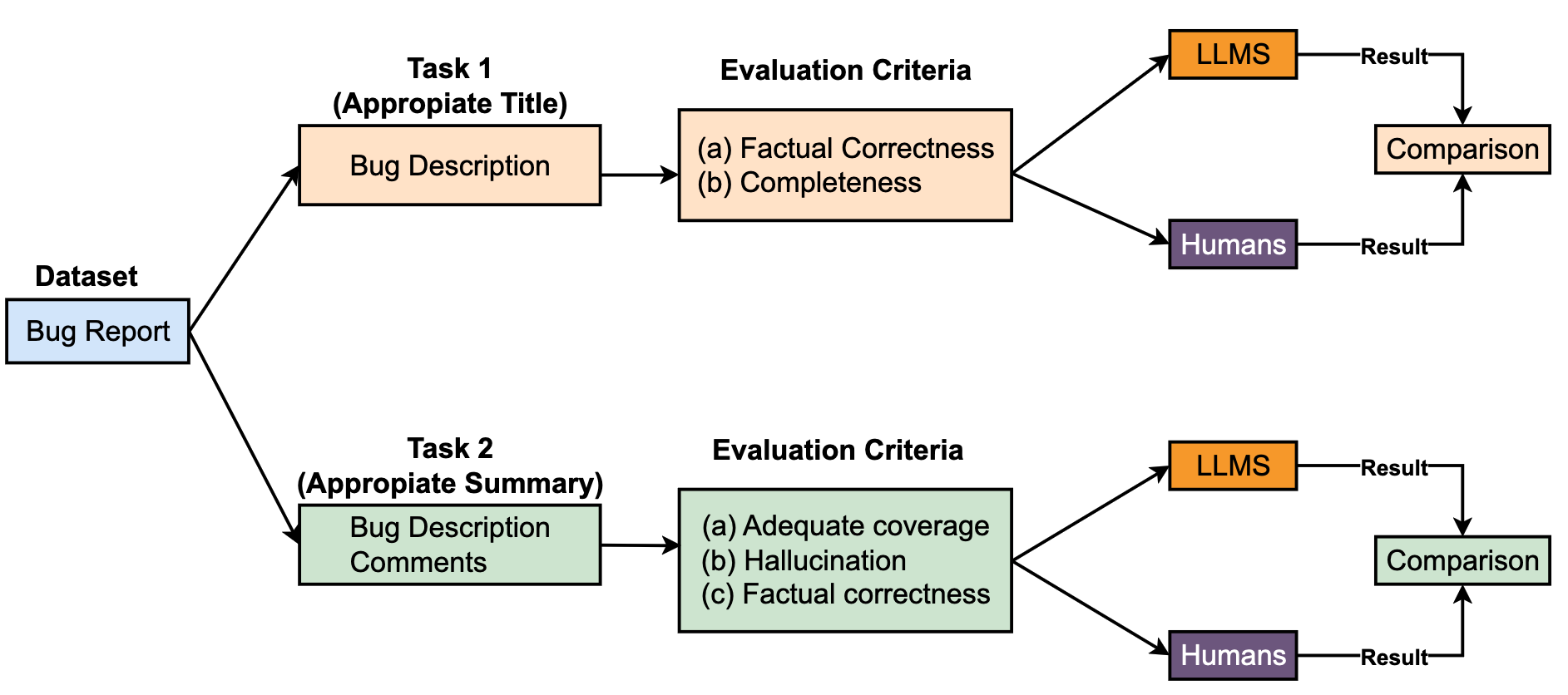}
	\caption{The overview of our study}
    \label{flow}
\end{figure}

Bug summarization condenses a bug report into a concise summary, highlighting essential information to help developers quickly understand and prioritize issues, facilitate efficient bug fixing, reduce debugging time, and improve software quality. Its importance is well recognized in the field \cite{br_1, br_2, br_3, br_4}. Given this, we investigated the potential of Large Language Models (LLMs) in evaluating bug report summaries by performing a study that compared the performance of GPT-4o \cite{gpt4}, LLaMA-3 \cite{llama3}, and Gemini \cite{gemini}, with human evaluators. This section describes all aspects of the design of the study and Figure \ref{flow} shows an overview of the study.

\subsection{Dataset Preparation}

We used the Signal repository hosted on GitHub, which contains a large collection of bug reports. We selected a subset of these reports for our study, ensuring that they were diverse and representative of real-world bug reports. Specifically, we chose 20 bug reports for Task 1, which involved evaluating bug titles, and 10 bug reports for Task 2, which involved evaluating bug report summaries (based on bug description and comments). 


\subsection{Task Description and Evaluation Criteria}

\subsubsection{Task Description}

In Task 1, we asked participants to choose the most appropriate bug report title from four options based on a given bug description. The correct option (title) was extracted from the GitHub repository, and we created the other three incorrect options for every question. There were 20 questions in total, divided into three difficulty levels: 8 \textit{easy} questions, 8 \textit{hard} questions, and 4 \textit{NOTA} (None Of The Above) questions, all randomly distributed in the questionnaire set. \textit{Easy} questions were very straightforward to answer, because three of the provided options were clearly incorrect, leaving the correct one to be the obvious choice. For questions of \textit{hard} difficulty, the options contained also an incorrect answer that was similar to the correct one, making the choice less obvious. We considered \textit{NOTA} questions the ones where the correct option was ``None of the above" since we believe that they require more effort and careful consideration of each option to determine that there was no correct answer.  We instructed participants to evaluate the titles on two criteria: factual correctness and completeness. We provided clear instructions and definitions of the evaluation metrics, along with examples and explanations to ensure participants understood the criteria. Although we cannot include these definitions and examples in the paper due to page limitations, they can all be found in our replication package \url{https://bit.ly/3zk7qZr}.

In Task 2, we asked participants to choose the most appropriate bug report summary from a set of options based on a given bug description and associated comments. There were 10 questions in total, with 6 being of \textit{hard} difficulty, 2 being \textit{NOTA}, and 2 \textit{easy}. Similarly to Task 1, \textit{NOTA} questions were those where the correct option was ``None of the above". We created all four options, including the correct option, based on the given bug description and associated comments. We instructed participants to evaluate the summaries based on three criteria: factual correctness, no hallucinations, and adequate coverage. The correct option met all these criteria, while the other options failed in one or more of these criteria. Again, we provided clear instructions and definitions of the evaluation metrics, along with examples and explanations to ensure participants understood the criteria. We provide these in our replication package.

Both tasks were conducted through Google Forms and participants were asked to evaluate each option based on the relevant criteria, using a scale of 0 to 2 for each metric. After evaluating each option, participants were asked to choose the title (for Task 1) or summary (for Task 2) they preferred, with the option to select ``None of the above" if they felt none of the options were correct or fulfilled the requirements.


We provided the same set of instructions, definitions, and questions to all three LLMs to ensure a fair comparison. Specifically, we instructed each LLM to evaluate the title options (for Task 1) and summary options (for Task 2) based on the same criteria used by human evaluators: factual correctness, completeness, and adequate coverage. By doing so, we aimed to eliminate any potential variability that could arise from differences in task interpretation. This approach allowed us to accurately assess the performance of each LLM and compare their results with those of human evaluators.

The complete data including the instructions, evaluation criteria, and question format for both tasks, can be found online \url{https://bit.ly/3zk7qZr}.

\subsubsection{Evaluation Metric}

For the evaluation of the responses from humans and LLMs, we used accuracy as the metric. Accuracy was calculated by comparing the evaluators' selections to the correct answers for each question. The accuracy for each question was calculated using the following equation:
\[
\text{Accuracy} = \left( \frac{\text{Number of correct evaluations}}{\text{Total number of evaluations}} \right) \times 100
\]

This method allowed us to account for the performance of each evaluator and provided a comprehensive measure of accuracy across different tasks and difficulty levels.



 \subsubsection{Large Language Model Selection}

 We have chosen three Large Language Models (LLMs)— GPT-4 \cite{gpt4} , Llama-3 \cite{llama3} and Gemini \cite{gemini} for our evaluation. These models were selected due to their proven effectiveness in handling programming-related problems and their training on extensive code bases. GPT-4, Llama-3, and Gemini have demonstrated strong performance in tasks such as code generation, bug detection, and understanding software-related text. Their ability to adapt to various areas of software development with high accuracy makes them suitable for assessing bug report summaries in our study.

\section{Results}

In this section, we present the results of our study, structured around four research questions (RQs). But first, we present the demographics of the study participants. 

\emph{Human Evaluators Demographics: }For Task 1, we had 13 evaluators, which answered all questions. The gender distribution comprised 9 male participants and 4 female participants. The participants' programming experience varied, with 6 having less than 5 years of experience, 4 with 5 to 10 years, and 3 with more than 10 years of programming experience. Regarding occupation, our participants were graduate students (3 participants), undergraduate students (3 participants), research scholars (4 participants), and professional software developers (3 participants).

For the Task 2, we had 10 evaluators. The gender distribution comprised 6 male participants and 4 female participants. The programming experience of the participants varied, with 5 having less than 5 years of experience, 3 with 5 to 10 years, and 2 with more than 10 years of experience. Regarding occupation, there were 3 undergraduate students, 2 graduate students, 2 research scholars, and 3 professional software developers.

\textit{RQ1: How did LLMs and humans perform in evaluating both tasks?}

\begin{figure}[!htbp]
    \centering
    \myfigureshrinker
    \begin{subfigure}[b]{0.49\columnwidth}
        \includegraphics[width=\textwidth]{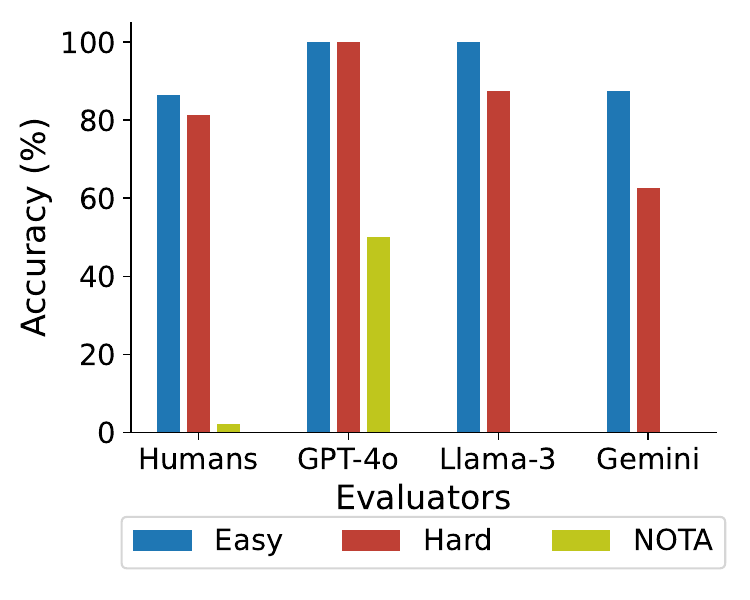}
        \caption{Task 1 (Titles)}
        \label{Overall_task1}
    \end{subfigure}
    \begin{subfigure}[b]{0.49\columnwidth}
        \includegraphics[width=\textwidth]{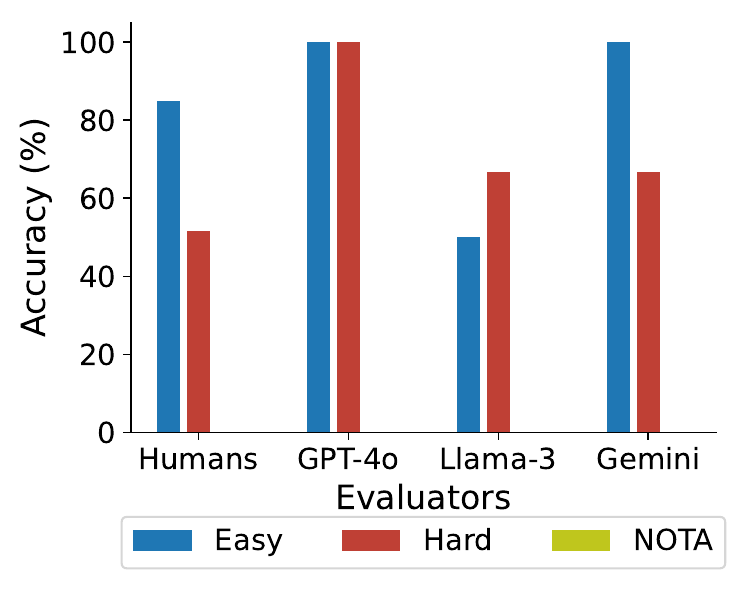}
        \caption{Task 2 (Summaries)}
        \label{Overall_task2}
    \end{subfigure}
    \caption{Comparison of Human Evaluators' and LLMs' accuracy in evaluating bug titles and bug summaries.}
    \label{fig:otasks}
\end{figure}

In Task 1, which involved evaluating bug titles, the results demonstrate a clear trend in performance between humans and the LLMs across varying levels of difficulty. On the easier tasks, GPT-4o and Llama-3 consistently performed at a high level, often outperforming human evaluators, particularly in cases where human judgment showed slight variability. As the difficulty increased to medium, the performance gap between GPT-4o and humans widened, with GPT-4o maintaining near-perfect accuracy. However, Llama-3's performance began to decline, indicating a sensitivity to the increased complexity of the task. On the NOTA tasks, all evaluators, including humans, struggled, but GPT-4o managed to maintain a notable performance, albeit reduced, compared to the easier tasks. In contrast, humans and the other LLMs exhibited a significant drop in accuracy, suggesting that these tasks were particularly challenging for non-expert systems and human evaluators alike. The results are shown in Figure \ref{Overall_task1}

In Task 2, which focused on evaluating bug report summaries, the overall trend remained similar to Task 1, with GPT-4o again leading in performance across all difficulty levels. On easy tasks, GPT-4o and Gemini matched each other, both demonstrating high accuracy, while humans showed a slightly lower performance, particularly as tasks became more complex. The medium difficulty tasks revealed a noticeable drop in human performance, likely due to the cognitive load and the intricacies involved in summarizing more complex bug reports. Llama-3 showed a moderate performance, but unlike GPT-4o, its accuracy fluctuated more noticeably. Interestingly, all evaluators, including GPT-4o, struggled with the NOTA tasks, where the evaluation of summaries became extremely challenging, leading to a uniform drop in performance. This indicates that while LLMs show promise in evaluating software summaries, there are still significant challenges in accurately assessing more complex or nuanced tasks. The results are shown in Figure \ref{Overall_task2}.

\textit{RQ2: How accurately do humans and LLMs assess factual correctness when evaluating bug titles and summaries?}

\begin{figure}[!htbp]
    \centering
    \myfigureshrinker
    \begin{subfigure}[b]{0.49\columnwidth}
        \includegraphics[width=\textwidth]{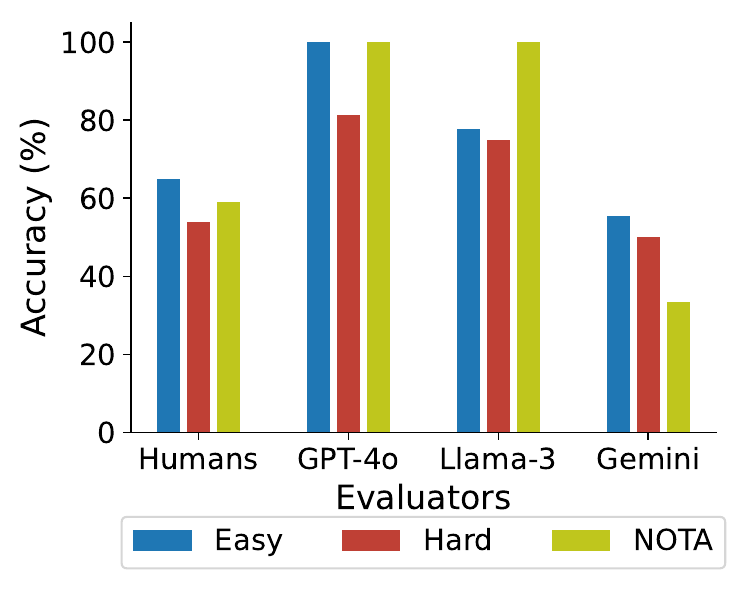}
        \caption{Task 1 (Titles)}
        \label{factual_task1}
    \end{subfigure}
    \begin{subfigure}[b]{0.49\columnwidth}
        \includegraphics[width=\textwidth]{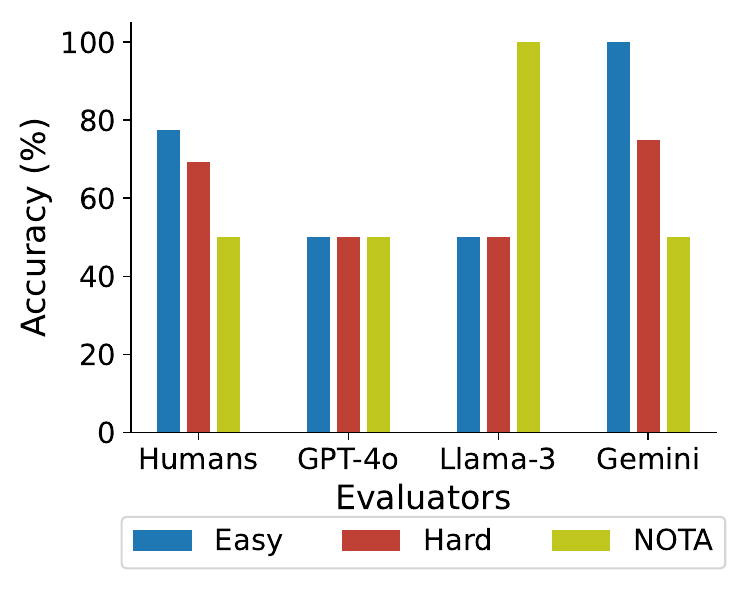}
        \caption{Task 2 (Summaries)}
        \label{factual_task2}
    \end{subfigure}
    \caption{Performance of Human Evaluators and LLMs in assessing factual correctness of the summaries}
    \label{fig:tasks}
\end{figure}

In Task 1, which involved evaluating the accuracy of factual information in bug titles, GPT-4o consistently outperformed the other evaluators across all levels of difficulty. Particularly on the easy and NOTA tasks, GPT-4o showed remarkable accuracy, suggesting that it handles factual data reliably, even in more challenging contexts. Humans, while performing well on the easy tasks, exhibited a notable drop in accuracy as the task difficulty increased, especially on the medium tasks. Llama-3 showed moderate performance across all levels, demonstrating some consistency but not matching the accuracy of GPT-4o. Gemini's performance was inconsistent, particularly struggling with medium and NOTA tasks, indicating potential challenges in processing factual data accurately in these contexts. The results are shown in Figure \ref{factual_task1}.

In Task 2, which focused on evaluating the factual content of bug report summaries, the results showed a different trend. While GPT-4o remained competitive, its performance was more variable compared to Task 1, particularly on easier tasks where it underperformed relative to the others. Humans, on the other hand, maintained relatively high accuracy across all levels of difficulty, with a slight drop on the NOTA tasks, reflecting their ability to consistently process and evaluate factual information. Llama-3 demonstrated a significant improvement on the NOTA tasks, outperforming both GPT-4o and Gemini, indicating its strength in handling complex factual evaluations. Gemini, while excelling in easier tasks, struggled with medium and NOTA tasks, further highlighting the variability in its performance when dealing with factual data. The results are shown in Figure \ref{factual_task2}.

\textit{RQ3: How effectively do humans and LLMs evaluate the completeness of bug titles and summaries?}

\begin{figure}[!htbp]
    \centering
    \myfigureshrinker
    \begin{subfigure}[b]{0.49\columnwidth}
        \includegraphics[width=\textwidth]{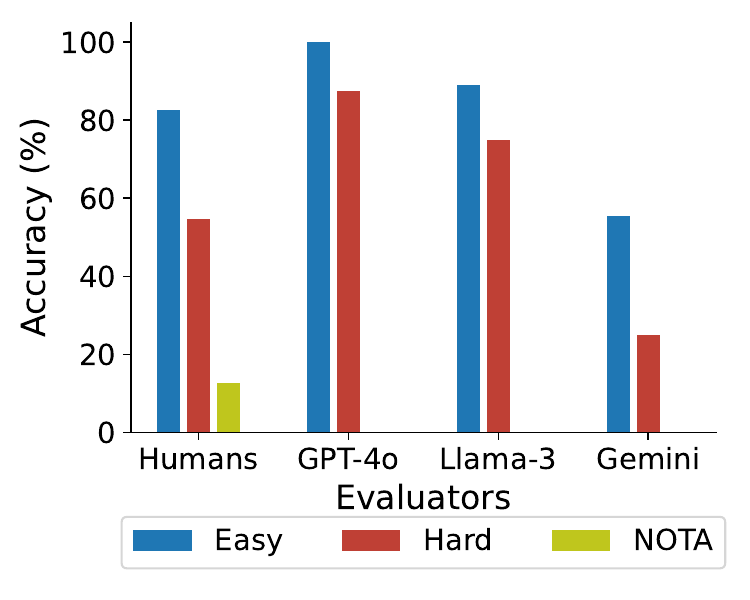}
        \caption{Task 1 (Titles)}
        \label{completeness_task1}
    \end{subfigure}
    \begin{subfigure}[b]{0.49\columnwidth}
        \includegraphics[width=\textwidth]{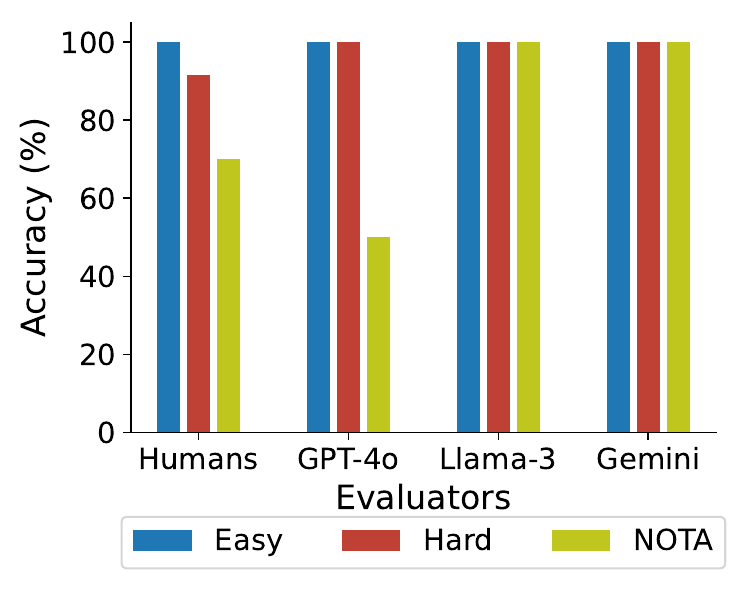}
        \caption{Task 2 (Summaries)}
        \label{completeness_task2}
    \end{subfigure}
    \caption{Performance of Human Evaluators and LLMs in assessing completeness of the summary.}
    \label{fig:tasks}
\end{figure}

In Task 1, which evaluated how thoroughly bug titles encapsulated all relevant information, GPT-4o consistently demonstrated strong performance, particularly on the easier tasks where it achieved perfect scores. Humans also performed well on the easier tasks but exhibited a significant drop in accuracy as the task difficulty increased. This drop highlights the challenges humans face in consistently capturing all relevant details, especially under more complex scenarios. Llama-3 showed a strong performance on the easy tasks but struggled similarly to humans as the difficulty increased. Interestingly, all evaluators, including humans, showed a significant decline on the NOTA tasks, with none of the LLMs, except GPT-4o, being able to capture the completeness of the titles accurately. This suggests that while LLMs can handle straightforward cases effectively, they may struggle with more nuanced or complex bug titles that require a deeper understanding of the context. The results are shown in Figure \ref{completeness_task1}.

In Task 2, which assessed the thoroughness of bug report summaries, showed a remarkable level of consistency across all evaluators, particularly on the easier tasks where all models, including humans, achieved perfect scores. As the task difficulty increased, GPT-4o and Llama-3 maintained their high performance, indicating their capability to thoroughly evaluate and understand the content of more complex bug reports. Humans showed a slight decrease in accuracy on the NOTA tasks, suggesting potential challenges in maintaining focus and attention to detail under more demanding conditions. Gemini also performed strongly across the board, particularly excelling in NOTA tasks, which suggests it may have an advantage in processing and summarizing more complex information. Overall, the results from Task 2 highlight that LLMs, particularly GPT-4o and Llama-3, are well-suited for evaluating the completeness of software summaries, potentially even outperforming human evaluators in more challenging scenarios. The results are shown in Figure \ref{completeness_task2}.

\textit{RQ4: How do humans and LLMs differ in their ability to identify hallucinations in bug summaries?}

\begin{figure}[!htbp]
    \centering
    \myfigureshrinker
    \includegraphics[width=0.3\textwidth]{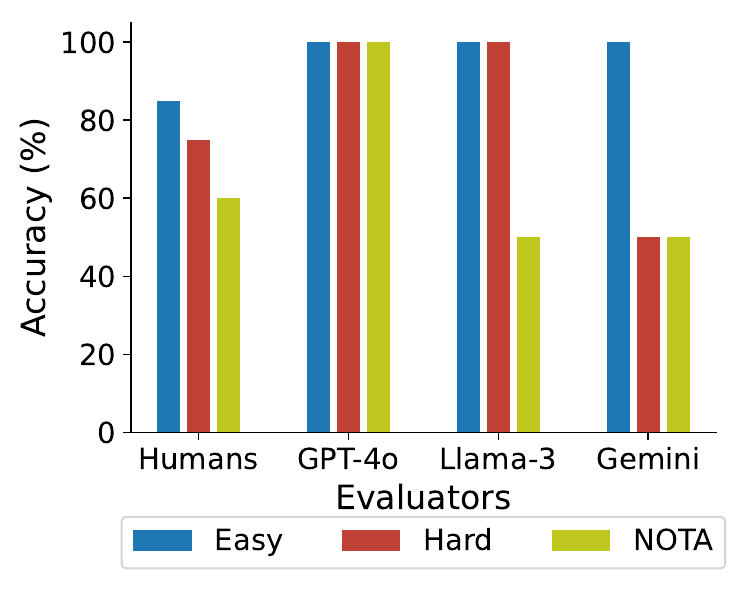}
    \caption{Performance of Human Evaluators and LLMs in detecting hallucination in the summaries.}
    \label{hallucination}
\end{figure}

In Task 2, we had one parameter, hallucination, which evaluated the propensity of the evaluators to avoid generating or selecting information that was not present in the original bug report summaries, GPT-4o consistently demonstrated exceptional performance across all levels of difficulty. Particularly on the easy and medium tasks, GPT-4o achieved perfect scores, indicating its strong ability to accurately interpret and evaluate the content without introducing errors or extraneous information. Llama-3 also performed well on the easier tasks but showed a significant drop in accuracy on the NOTA tasks, suggesting that while it is generally reliable, it may struggle with more complex summaries where the risk of hallucination increases. Humans, while performing well overall, exhibited a noticeable decline in accuracy as the difficulty level increased, highlighting the challenges they face in maintaining precision under more demanding conditions. Gemini showed high accuracy on easy tasks but struggled on both medium and NOTA tasks, particularly with a substantial drop on the NOTA tasks. This variability suggests that while Gemini can handle straightforward evaluations, it may be more prone to errors in more complex or ambiguous scenarios. Overall, GPT-4o stands out as the most reliable evaluator in this context, with minimal risk of hallucination across varying levels of task complexity.. The results are shown in Figure \ref{hallucination}.

\section{Conclusions and Future Work}

We conducted a preliminary study investigating whether LLMs can be used to evaluate issue report summaries.  Our findings show that LLMs performed remarkably well as evaluators across various evaluation criteria. While human evaluators excel in simpler evaluations, their performance tends to decline with increasing task complexity. In contrast, LLMs, particularly GPT-4o, demonstrate robust capabilities across different difficulty levels, indicating their potential as reliable tools in evaluating bug report summaries. Future work includes scaling up our participant numbers and diversity and studying LLMs as evaluators in a broader range of tasks. 


\bibliographystyle{IEEEtran}
\bibliography{ref}

\end{document}